\documentclass[sigconf,screen,authorversion,nonacm]{acmart} %%authorversion,nonacm

\usepackage{times}
\usepackage{url}
\usepackage{listings, lstautogobble}
\usepackage{xspace}

\usepackage[nameinlink]{cleveref}
\usepackage{hyperref}

\usepackage{balance}

\newcommand{\bosque}{\textsc{Bosque}\xspace}
\newcommand{\bapi}{\textsc{BAPI}\xspace}

\newcommand{\tecton}{\textsc{Tecton}\xspace}
\newcommand{\sundew}{\textsc{Sundew}\xspace}

\newcommand{\etc}{\hbox{\emph{etc.}}\xspace}
\newcommand{\vs}{\hbox{\emph{vs.}}\xspace}

\newcommand{\cf}[1]{\texttt{#1}}

\newcommand{\codefont}{\footnotesize}

\definecolor{cgreen}{rgb}{0.25,0.5,0.35} % comments
\definecolor{stringred}{rgb}{0.6,0,0} % for strings

\lstdefinelanguage{bosque}{
keywords={api, function, action, requires, ensures, env, permissions, example, type, \$events, \$return, agent, let, if, none, return},
keywordstyle=\color{blue}\bfseries,
identifierstyle=\color{black},
alsoother={@},
sensitive=true,
comment=[l]{\%\%},
commentstyle=\bfseries\color{cgreen}\ttfamily,
}
 
\setcopyright{cc}
\setcctype{by}
\acmDOI{10.1145/3842652.3843194}
\acmYear{2026}
\copyrightyear{2026}
\acmISBN{979-8-4007-2970-6/2026/10}
\acmConference[SpecOps '26]{Proceedings of the 1st International Workshop on Specification-Driven Development Life Cycle}{October 4--9, 2026}{Oakland, CA, USA}
\acmBooktitle{Proceedings of the 1st International Workshop on Specification-Driven Development Life Cycle (SpecOps '26), October 4--9, 2026, Oakland, CA, USA}
\acmSubmissionID{splashws26specopsmain-p3-p}
\received{2026-06-12}
\received[accepted]{2026-07-16}



\begin{document}

\title{Specifications for Humans, Agents, and Tooling}
\subtitle{\bosque API (BAPI) Tooling}

\author{Mark Marron}
\orcid{0000-0003-3589-4860}
\affiliation{%
  \institution{University of Kentucky}
  \city{Lexington}
  \country{USA}
}
\email{mark.marron@uky.edu}

\begin{abstract}
Specifications are the central mechanism for communicating intents, requirements, and constraints in software development. When they are explicit, 
clear, and reliable, they are an effective means for collaboration and cooperation. They allow for stakeholders 
to specify what they want, developers (or AI agents) to understand and implement the needed functionality, for clients to effectively use the system, 
and for automated tooling to validate the correctness for each of these steps. 

This tool paper outlines the \bosque API (\bapi) ecosystem, a software ecosystem designed to support modern \emph{spec-centered} development. The \bapi specification 
language works in a fully polyglot ecosystem and provides a suite of features, including unparalleled expressivity, test generation, validation, and sand-boxing to support the complete 
application development lifecycle. These are critical to supporting emerging security and coding (both API implementation \& usage) challenges presented by 
agentic AI systems. 
\end{abstract}

\maketitle

\begin{center}
\small
\textit{Authors copy -- final version to appear Proceedings of the 1st International Workshop on Specification-Driven Development Life Cycle (SpecOps '26)}
\end{center}

\vspace{0.5em}

\section{Introduction}
Software engineering is moving from monolithic codebases to modular and polyglot, API centric architectures. In some cases, the API centric transformation is simply a
mechanism for managing complexity but it also represents a new model of development focused on composition and integration of existing code or services. The appeal of API 
centric development is clear -- it enables code reuse and the reorganization of developers into agile teams, which, combined with API-unlocked re-use, enable developers to focus 
on key functionality and features. This shift places API contracts in a critical role. Inadequate contracts --- whether due to weak type systems, ambiguous 
documentation, or insufficient testing --- lead to mismatched expectations. These gaps can produce outcomes ranging from simple crashing bugs to 
severe security failures or data breaches. 

Closing these gaps via comprehensive specification is the foundation for ensuring that a software artifact matches the intent of the stakeholders, that it has been correctly 
implemented, and enables clients to use the system without making (risky) assumptions. Specifications become even more critical in the context of agentic software development.
Without appropriate structure and information agentic systems quickly start to make erroneous assumptions and generate incorrect code or take dangerous actions. 

\begin{figure}[t]
\centering
\vspace{5mm}
\begin{lstlisting}[language=bosque,basicstyle=\codefont\ttfamily]
%% Transfer amt from payer account to payee account.
api transfer(amt: USD, payer: Account, payee: Account);
\end{lstlisting}
\caption{Baseline specification for a (dangerous) payment API using only a type signature similar to what might be seen in TypeSpec~\cite{typespec}, Smithy~\cite{smithy}, OpenAPI~\cite{openapi} \etc} 
\label{fig:origapi}
\end{figure}

Given the risk of financial loss due to mis-use, or malicious attack, around the API in \Cref{fig:origapi}, would you trust your AI agent to use the API below to make payments from your bank account? 
In fact even with human level intelligence (true AGI) this kind of API has proven\footnote{Notable incidents of this sort include~\cite{fatfinger,faterfinger} with direct financial losses 
exceeding $\$50$ million in addition to indirect reputational damage.} to be poorly specified and definitely unsafe! The API is missing critical information about how to use the API, 
what it can affect, and lacks any sort of safety measure to prevent undesired behavior or limit possible damage.

Addressing the challenges that software developers face with usability and correctness today along with the new challenges presented by agentic software development, the new risks of 
malicious prompt injection, data exposure, and reward hacking, requires a shift to an aggressively \emph{spec-centered} method of software engineering. In this approach, specifications are the 
primary artifact of the software development process, and from them, flows the implementation, testing, deployment, and interaction with AI agents, data, and other systems. This paper outlines 
the \bosque API (\bapi) specification ecosystem, consisting of the \bapi specification language, which supports specification of data, behavior, temporal relations, and resources. We also describe 
tooling to support validation and testing of implementations, automated polyglot runtime interoperation \& data quality assurance, along with specialized features for agentic development.

The \bapi specification language takes the current state-of-the-art for REST API specifications for inspiration, specifically 
TypeSpec~\cite{typespec} and Smithy~\cite{smithy}, and the core ideas that an API specification should be more akin to a programming language than simply a 
description of a data-transport or encoding schema. From this viewpoint the \bapi design uses a rich type system~\cite{bosque,bsqon} that provides 
a range of Poke-Yoke~\cite{pokeyoke} features to prevent common errors and a language that allows us to provide detailed logical operational constraints on the behavior of the API 
(\Cref{sec:bapi}). \bapi extends this specification layer with novel capabilities to express temporal constraints across API uses as well explicit resource access annotations 
(\Cref{sec:bapi,fig:finalapi}). Using this specification language we show how to build a rich ecosystem of tools for mechanized validation of implementations and their 
usage (\Cref{sec:tooling}). Finally, we discuss the implications of this design for agentic software development and how it can be used to support 
safe and effective agentic interactions (\Cref{sec:agentic-coworkers}).

\begin{figure}[t]
\centering
\begin{lstlisting}[language=bosque,basicstyle=\codefont\ttfamily]
%% Transfer amt from payer account to payee account.
api transfer(amt: USD, payer: Account, payee: Account)
  requires 0.0<USD> < amt;
  requires amt <= 100.00<USD>;
;
\end{lstlisting}
\caption{Basic \bapi specification for the payment API with constraints (only constants) on the transfer amount.} 
\label{fig:basicbosque}
\end{figure}

\section{\bapi Specification Language}
\label{sec:bapi}
The core of \bapi is a language (\bosque~\cite{bosque,bsqon}) that allows us to specify the structure of data in the style of a type system, instead of a data transport 
schema, which ensures that these specifications can be (type) checked, composed, versioned, just like any other software artifact. 

In this process we lean into the ideal 
of mistake-proofing the specification process~\cite{pokeyoke}. An example of this philosophy in physical hardware is the USB-A \vs USB-C design, where the symmetrical design 
of the C connector mistake-proofs the process of connecting a cable. In the context of safety one can look at the use of \emph{lockout/tagout}~\cite{loto} procedures in industrial 
settings where each worker uses a unique and personal lock and tag to make it impossible for anyone (or anything) to activate a system until all workers have released their 
locks. In the context of API design we can look at the ability to make an API denial-of-service mistake-hardened by providing explicit syntax and language support for 
\emph{regex validated string} types -- \cf{type UserID = CString of /[a-zA-Z0-9\_]\{2, 30\}/} -- and guaranteeing that the regex execution cannot be used to trigger a denial-of-service 
attack (ReDOS~\cite{redos})\footnote{Unfortunately, the OpenAPI, TypeSpec, and Smithy \cf{pattern} annotations use JavaScript regexes 
which can trigger ReDoS attacks. Thus, these specifications can provide a very convenient data-source~\cite{redossearch} of ReDoS vulnerabilities for attackers to mine.}. 

From the core \bosque type language, the next step is to support arbitrary logical constraints on data relations and behavior of the APIs.
In \bapi this is done by allowing for invariant declarations on types and pre/post conditions on API operations. However, instead of directly using a dialect of first-order logic, 
\bapi leverages the \bosque programming language~\cite{bosque}. 

\bosque is uniquely well suited for this task as it is specifically designed 
to support mechanized validation and, as such is fully referentially transparent and deterministic, which eliminate many of the challenges around accidental mutation or nondeterministic 
failures when trying to use a general purpose programming language for specification and runtime validation~\cite{dafny,specsharp,iivv}. In addition, the \bosque language is designed to map well into 
the families of logics that are efficiently handled by SMT~\cite{z3} solvers.

Using this logical specification layer we can improve the payment API example by adding basic sanity check pre/post conditions on the operation as shown in \Cref{fig:basicbosque}.

This specification now provides a much safer API, as it prevents the risk of simple fat-finger mistakes or malicious attempts to subvert an AI agent into transferring large amounts of money. 
In addition this specification also highlights (and prevents) the possible error of using a negative amount for the transfer request. However, this API specification is still unsatisfactory 
as the upper limit on the transfer amount is a hard coded constant and thus not flexible to changes in the business rules or user needs. This specification is also silent on the possible side-effects 
of the API operation -- for example does it require authorization and access permissions for the payee account, and, is it possible that this API also uploads my account information to 
\emph{hackers.evil.com}?

To address these issues \bapi provides two forms of \emph{environmental constraints} that can be used in the specification language. The first constraint is a set of environment variables 
that are required for the API to function, such as a \cf{PAYMENT\_AUTHORIZATION} token and the user specified \cf{PAYMENT\_LIMIT}. The second constraint is a set of resource access annotations 
that specify the resources that the API will access, in this case \emph{just} the payer’s account information. Explicitly declaring these requirements ensures that a user (human or agent)
has the necessary context and also, depending on the language/runtime used to implement the API\footnote{A fully managed runtime like \bosque can strongly enforce these constraints while other 
implementation platforms, like C on Linux, may need to use VM or hypervisors to enforce these constraints or may only provide partial guarantees. However, from our standpoint the ability to express these 
resource attributes is a key feature for enabling any form of secure computation.}, can sandbox access to only information that is explicitly provided. 
Adding these constraints to the API specification we get the enhanced specification shown in \Cref{fig:enhancedbosque}.

\begin{figure}[t]
\begin{lstlisting}[language=bosque,basicstyle=\codefont\ttfamily]
%% Transfer amt from payer account to payee account.
api transfer(amt: USD, payer: Account, payee: Account)
  env={
    PAYMENT_AUTHORIZATION: OAUTH_TOKEN,
    PAYMENT_LIMIT: USD
  }
  permissions={
    \account:${payer.routing}/${payer.account}\
  }

  requires 0.0<USD> < amt;
  requires amt <= env.PAYMENT_LIMIT;
;
\end{lstlisting}
\caption{Enhanced \bapi specification for the payment API with environment variable requirements and resource access annotations.}
\label{fig:enhancedbosque}
\end{figure}

The upper bound on the transfer amount is now determined by the user specified \cf{PAYMENT\_LIMIT} environment variable which allows for dynamic adjustment based on the context that 
this call is made in. In addition, the explicit resource \cf{permissions} annotation provides a "sandbox" specification for the resources that this API will access. While the idea 
of sandboxing is well known, the key insight here is using URIs and Globs~\cite{globs} as the means of resource specification and sandboxing. As opposed to using custom resource 
types and access languages~\cite{javaresource,bsdjails,ztdjava} which have, historically, been difficult to standardize, understand, and use, URIs and globs naturally match the model 
of resources used in RESTful systems. This allows for a uniform and natural way to specify and enforce access policies across a wide variety of resource types 
(files, REST endpoints, databases, \etc) without needing custom plugins or extensions for each type. 

These environmental constraints are critical for ensuring that the API is safe to use and that it has the necessary context to function correctly. However, we still lack the critical 
ability to express temporal constraints on the API operations. For example, that after a blob storage write that a conditional (\cf{If-Match}) e-tag read will, on success, have the same value 
is the same as previously written, or in the payment API example, that either the payment amount does not exceed a certain limit or that in some previous event we obtained explicit user approval 
for the operation.

\bapi provides a novel event log system to express temporal constraints on the API operations. This tamper-proof log is a first-class construct in the specification language and allows 
us to write specifications over the sequence of events that occur during the execution of a system. In contrast to classical temporal logics~\cite{ltl,ctl} with 
limitations on expressiveness\footnote{Consider any non-regular property, such as the number of cloud VM provision calls is equal to the number of release events, which cannot be expressed 
in classical temporal logics but are expressible in a standard way writing functions over the \cf{event log} list contents.} or TLA~\cite{tla} which is not executable, this approach allows 
developers to express complex temporal constraints on the API operations in a natural and intuitive way. 

Each event in the system is recorded in the event log in linear order, and as part of the pre or post conditions of an API operation, this event log is automatically provided via the special 
variable \cf{\$events} that provides the standard set of List type operations plus special pattern matching query operations. This list is tamper-proof in that only the runtime may add entries 
to the list and does so based on the actual results of any API calls that are made. This prevents accidental, malicious, or reward hacking to forge parts of an event log to satisfy a precondition, 
such as making up a customer ID instead of loading it from the appropriate database. 

Using this log, and a set of seqential list style data processing operations, we can express a rich set of temporal constraints on the API operations. In the payment API we can add a temporal 
constraint that allows for the transfer amount to exceed the default payment limit \emph{but only} when a user has explicitly approved this payment\footnote{Note this spec is still incomplete 
as it will still allow double payment or an unapproved transfer that happened to match the amount.}. The code in \Cref{fig:finalapi} uses a 
\cf{contains} check to search the event log for a record of user approval, prior to using this API, if it exceeds the default payment limit. As this check is entirely over locally observed 
activity and does not involve re-querying any external (global) state, which may change between the time of the approval and the time of the transfer, we avoid the potential for time-of-check 
time-of-use (TOCTOU) races.

\begin{figure}[t]
\centering
\begin{lstlisting}[language=bosque,basicstyle=\codefont\ttfamily]
%% Transfer amt from payer account to payee account.
api transfer(amt: USD, payer: Account, payee: Account)
  env={
    PAYMENT_AUTHORIZATION: OAUTH_TOKEN,
    PAYMENT_LIMIT: USD
  }
  permissions={
    \account:${payer.routing}/${payer.account}\
  }

  requires 0.0<USD> < amt;
  requires amt <= env.PAYMENT_LIMIT ||
    $events.contains(Approve{|payee=payee, amt=amt|});
;
\end{lstlisting}
\caption{Complete \bapi specification for the payment API -- including logical constraints, resource access permissions, and temporal constraints.} 
\label{fig:finalapi}
\end{figure}

\section{Tooling Ecosystem}
\label{sec:tooling}
In the vision of a \emph{spec-centered} development ecosystem, specifications drive the development, testing, and deployment\footnote{\bapi can be used to generate canonical OpenAPI specifications for integration with 
existing systems and deployment/operations tooling.} of the system. The \bapi ecosystem includes high-value tools and workflows that 
support this vision. 

\subsection{Automated Test Generation}
Test generation is a critical component of the software development process, and arguably, one of the less enjoyable and more taxing tasks for developers. However, using the 
rich type and specification information provided by \bapi, the ability of LLM's to generate contextually-relevant small values, and the fault-exposing capabilities of 
combinatorial testing~\cite{combtesting} we can generate effective test suites for validating the correctness of implementations and uses of APIs. 

The \tecton~\cite{tecton} tool is an automated test generation system that, given only a \bapi specification and (optionally) any existing mock data-sources, generates a 
comprehensive test suite\footnote{Note that \tecton is entirely agnostic to the language used to implement an API and, in fact, the implementation does not 
even need to exist!}. On a set of standard RESTful testing benchmarks~\cite{tecton,arat-rl,autoresttest} \tecton achieves an effective rate of $70\%-90\%$ line coverage, 
comparable to quality human generated suites, and is able to expose between $2\times$ and $4\times$ the number of faults as a state-of-the-art white-box fuzzing 
systems~\cite{autoresttest,evomaster,restler}. Further, given the context (token) efficient nature of \bapi specification and the ability to focus the LLM on compact tasks, 
these test suites can be generated for under USD $\$0.10$ in LLM inference cost.

\subsection{Automated Validation with \sundew}
In cases where an API implementation or an API client is \emph{implemented in} the \bosque programming language we can go beyond simply testing and apply formal mechanized validation 
to check that the implementation or use is correct with respect to the specification. As described by the \bosque developers~\cite{formalmorgan}, the design of \bosque enables us 
to map it, almost entirely, to efficiently decidable theories supported by SAT-Module-Theory (SMT) solvers~\cite{z3}. Operations on numbers, data-types, and functions all map to core 
decidable theories -- Integers, Bitvectors, Constructors, Uninterpreted Functions, and Interpreted Functions. In SMT solvers the theories of Strings and Sequences 
are semi-decision procedures in the unbounded case.

\sundew performs small-model validation by bounding the sizes of the String and Sequence values, leveraging the small-model hypothesis that
the vast majority of errors can be triggered by small (bounded) inputs, and transforming the semi-decision procedure into a fully (and efficiently) decidable system. As a result validation 
can be fully automated and the result is either (1) a proof that no small input exists that can trigger the error, or (2) a counterexample input that does trigger the error. There is no 
need for developers to learn an additional proof language, write additional proof lemmas or tactics, or understand the nuances of the underlying logic engine to debug proof failures. 

In the case where a full proof-of-correctness is desired, it is possible to use standard verification-condition and AI generated inductive invariants~\cite{lean,dafny,fstar} to perform 
modular full-verification of a \bosque program. As \bosque code is the same as the specification logic language this process is applicable to any property that is part of a \bapi 
specification -- or part of an agentic task (\Cref{sec:trustworthy-agents}).

\section{Agentic Coworkers}
\label{sec:agentic-coworkers}
The rise of agentic software systems presents a new set of challenges for software development and specification systems have a key role to play in this space. In order for an AI 
system or Agent to be trustworthy, beyond a statistical guarantee that it "usually does the right thing", we must have a clear and unambiguous specification of the system behaviors 
it interacts with and the ability to validate its behavior with respect to these requirements. \bapi includes two important features
in this space -- explicit support for multi-modal specification for AI coding assistants and support for mechanized validation of agentic interactions.

\subsection{Multi-Modal Specification for AI Coding Assistants}
Effective specifications for agentic software development need to be multi-modal including natural language, 
examples, and logical constraints. \bapi specifications, as described in \Cref{sec:bapi}, cover all of these modalities except for examples. While, from a theoretical perspective, 
examples (or test cases) are not strictly necessary, as they can be subsumed by sufficiently strong logical constraints, they are critical in practice for communicating intent concisely 
when a comprehensive logical specification is difficult to write or understand and can effectively eliminate ambiguity in many cases~\cite{intentformal,uist,flashfill}.

To address this issue \bapi includes an \cf{example} keyword for declaring example-based specifications that allows developers to provide inline, in a file, or as a directory of 
examples that can be used as tests for validating the correctness of an implementation or given to an AI agent as part of the context for generating code. Consider the classic 
case of an API for a sorting routine that sorts a list of integers. A common mistake in the spec it to only write the postcondition that the output is sorted but neglect to 
specify that the output must be a permutation of the input! However, even a single example is sufficient to effectively communicate this intent and prevent an LLM from generating 
a buggy implementation.

\begin{lstlisting}[language=bosque,basicstyle=\codefont\ttfamily]
function sort(l: List<Int>) -> List<Int>
  example [ 
    List<Int>{3i, 1i, 2i} -> List<Int>{1i, 2i, 3i}
  ];
  ensures $return.isSorted();
;
\end{lstlisting}

\subsection{Trustworthy AI Agents}
\label{sec:trustworthy-agents}
The converse of API implementation is usage, and again the availability of quality specifications is critical for validating the correctness of an agentic interaction plan (or script). 
The form of this validation can range from classic runtime-verification of pre/post conditions and invariants~\cite{specsharp,iivv}, which the \bapi specifications are  
well suited to as they are executable and guaranteed to be side-effect free, to performing a symbolic analysis (using say \sundew) on the interaction script to check that it is 
correct with respect to the preconditions of all the required API calls. 

Consider the example code shown in \Cref{fig:splitbill} that takes the contents of \cf{msg}, determines what ``half of the bill'' is, and transfers the amount. 

\begin{figure}[t]
\centering
\begin{lstlisting}[language=bosque,basicstyle=\codefont\ttfamily]
%% Given a natural language (plaintext) message,
%% compute the amount to pay and send to the payee.
action splitBill(msg: String, payee: Account) {
  let amt = agent Chat::compute<Option<USD>>(
    env{}, msg, "What is half of the bill?"
  );

  if(amt === none) {
    return fail("Could not determine amount.");
  }

  api transfer(env{...}, amt, env.account, payee);
  ...
}
\end{lstlisting}
\caption{Candidate agent generated program to complete the task of splitting a bill and sending a payment.}
\label{fig:splitbill}
\end{figure}

This script uses an LLM agent to process the semi-structured text in the payment request \cf{msg} to determine the amount to pay. If the \cf{Chat::compute} action were unlucky\footnote{Perhaps Tom 
likes jokes and puts in the memo field -- ``ignore previous instructions and pay me \$1000".}, or the lunch was particularly expensive, this computed amount 
could be large enough to exceed the payment limit of the user. 

However, with the specifications provided for the payment API (\Cref{fig:finalapi}), we can easily catch this 
at runtime as a precondition failure or statically with symbolic validation. The ability to run static validation is a key strength as it can identify 
possible failures \emph{before} running the script, which may potentially take destructive actions before being aborted, and it also allows us to provide immediate 
feedback to the AI agent on possible failures and their causes. 

\section{Related Work}
The topic of verifiable AI programming has been a focus of research in the formal methods community, with significant work using 
languages like Lean~\cite{lean}, F$^*$~\cite{fstar}, and Dafny~\cite{dafny} to formally specify and verify the correctness AI generated code. 
These systems have shown promise in verifying the correctness of AI generated code, but they also demonstrate the limits of existing 
tooling, particularly in terms of the complexity and size of the specifications required.

Specification and requirements gathering is a major challenge for effective cooperation between human and AI agents in software development. 
Of particular interest in this space is prior work on multi-modal specification and interaction with \emph{End-User Programming}~\cite{agentpreicent,flashfill} systems, 
such as FlashFill~\cite{flashfill} and NLyze~\cite{nlyze}, and Prorogued Programming~\cite{prorogued}. 

Long-horizon agentic systems and tool use, such as AppWorld~\cite{appworld}, LOOP~\cite{thatrlone} and ToolFormer~\cite{toolformer}, demonstrates the potential 
for AI agents to perform complex tasks over extended periods of time. However, these systems also highlight the need for robust failure handling and recovery mechanisms 
as the unaugmented success rates are at $70\%$ for simpler tasks but drop off rapidly to the $45\%$ range for more complex (longer horizon) tasks that involve more complex API (tool) 
use. These results highlight the need for more robust specification and validation mechanisms to construct verifiably trustworthy agentic systems~\cite{agentictrust}.

\section{Onward!}
This paper outlines the Bosque API (\bapi) ecosystem\footnote{All of the systems described in this paper are actively developed and publicly available via 
\url{https://github.com/BosqueLanguage/BosqueCore}.}, a software ecosystem designed to support modern \emph{spec-centered} development and 
provides examples of high-value tools and workflows that support this vision including automated test generation, support for agentic code generation, and 
mechanized validation. 
We believe \emph{spec-centered} software development, and \bapi, will play a critical role in addressing existing challenges in the software engineering lifecycle as 
well as emerging security and coding challenges presented by agentic systems. 

\begin{acks}
    I would like to thank the anonymous reviewers for their feedback and suggestions on this work. I would also like to thank Stephen Goldbaum for early discussions and ideas on 
    how developers can specify and validate system behaviors as entirely local observations, which was a key insight in the design of the \bapi event log system. I would like 
    to thank Brian Terlson for sharing his experiences with API design and specification and the TypeSpec project along with Richard Perris and Claudia Babescu for letting me 
    work with them on a great spec-centered project. 
\end{acks}

\balance
\bibliographystyle{ACM-Reference-Format}
\bibliography{bibfile}

%\section*{Appendix}
%\label{sec:appendix}

\end{document}